\documentstyle[prl,aps]{revtex}

\draft
\newcommand{\ba}{\begin{eqnarray}}
\newcommand{\ea}{\end{eqnarray}}
\input{psfig}

\begin{document}
\title{Cluster states in nuclei as representations of a $U(\nu +1)$ group}
\author{R. Bijker$^1$ and F. Iachello$^{2}$}
\address{$^1$Instituto de Ciencias Nucleares, 
Universidad Nacional Aut\'onoma de M\'exico, \\ 
Apartado Postal 70-543, 04510 M\'exico, D.F., M\'exico \\
$^2$ Center for Theoretical Physics, Sloane Laboratory, \\
Yale University, New Haven, CT 06520-8120, U.S.A}
\date{January 27, 2000}
\maketitle

\begin{abstract}
We propose a description of cluster states in nuclei in terms of
representations of unitary algebras $U(\nu +1)$, where $\nu $ is 
the number of space degrees of freedom. Within this framework, a 
variety of situations including both vibrational and rotational 
spectra, soft and rigid configurations, identical and non-identical 
constituents can be described. As an example, we show how the method 
can be used to study $\alpha$ clustering configurations in $^{12}$C 
with point group symmetry ${\cal D}_{3h}$. 
\end{abstract}


The purpose of this letter is to point out that the algebra $U(\nu +1)$,
which has been suggested to be the spectrum generating algebra for a quantum
mechanical problem with $\nu $ space degrees of freedom \cite{iac}, might
provide a framework for a unified description of cluster states in nuclei.
We note that the main properties of clustering in nuclei are: (i) the
softness of the cluster configuration which makes nuclei appear more like
liquid structures rather than rigid molecular structures in which the
constituents sit at some definite location in space; (ii) the near equality
of vibrational and rotational energies which does not allow a clear-cut
distinction between these two types of motion; (iii) the fact that the
constituents are not point-like objects but particles with a spatial extent
comparable to that of the overall structure and (iv) the fact that the
constituents are often identical which implies that permutation symmetry
must be imposed. A unified description of clustering in nuclei should be
able to accomodate all these properties.

To illustrate the uselfuness of the algebra $U(\nu+1)$ in describing the
variety of observed situations, we consider the specific case of a cluster
composed of three particles (a description of two-body
cluster configurations in nuclei in terms of $U(4)$ was suggested long ago 
\cite{iac1} and has been used to describe resonances in heavy ion scattering 
\cite{bro}). For a three-body problem, the number of space degrees of
freedom (after removal of the center of mass) is $\nu =3n-3=6$. (We do not
consider in this article constituents with an internal structure. For such
situation the algebraic structure must be enlarged to $U(\nu+1) \otimes 
U(\Omega)$ where $\Omega$ is the number of internal degrees of freedom.)
The space degrees of freedom can be taken as the Jacobi coordinates $\vec{%
\rho}=(\vec{r}_{1}-\vec{r}_{2})/\sqrt{2}$ and $\vec{\lambda}=(\vec{r}_{1}+%
\vec{r}_{2}-2\vec{r}_{3})/\sqrt{6}$, where $\vec{r}_{i}$ $(i=1,2,3)$ are the
coordinates of the three particles. The corresponding algebra is $U(7)$. The
algebra of $U(7)$ is constructed by introducing two vector bosons $b_{\rho }$%
, $b_{\lambda }$ together with an auxiliary scalar boson $s$. It was
introduced in \cite{bij} where it was used to describe three-quark
configurations in baryons. The 49 bilinear products of creation and
annihilation operators generate the Lie algebra $U(7)$, 
\begin{eqnarray}
b_{\rho ,m}^{\dagger }~,\;b_{\lambda ,m}^{\dagger }~,\;s^{\dagger } &\equiv
&c_{\alpha }^{\dagger }\hspace{1cm}(m=0,\pm 1)\hspace{1cm}(\alpha =1,\dots
,7)  \nonumber \\
{\cal G} &:&G_{\alpha \beta }\;=\;c_{\alpha }^{\dagger }c_{\beta }\hspace{1cm%
}(\alpha ,\beta =1,\ldots ,7)
\end{eqnarray}
The creation and annihilation operators for vector bosons 
($b_{\rho,m}^{\dagger}$, $b_{\lambda,m}^{\dagger}$ and 
$b_{\rho,m}$, $b_{\lambda,m}$) represent the second quantized form 
of the Jacobi coordinates and their canonically conjugate momenta, 
while the auxiliary scalar boson is introduced in order to
construct the spectrum generating algebra. (The method of embedding the
problem in a larger dimensional space \cite{iac} is similar to that used in
Kaluza-Klein theories of particle physics.) The energy levels can be
obtained by diagonalizing the Hamiltonian 
\begin{eqnarray}
H &=&H_{0}+\epsilon _{s}\,s^{\dagger }\tilde{s}-\epsilon _{p}\,(b_{\rho
}^{\dagger }\cdot \tilde{b}_{\rho }+b_{\lambda }^{\dagger }\cdot \tilde{b}%
_{\lambda })+u_{0}\,(s^{\dagger }s^{\dagger }\tilde{s}\tilde{s}%
)-u_{1}\,s^{\dagger }(b_{\rho }^{\dagger }\cdot \tilde{b}_{\rho }+b_{\lambda
}^{\dagger }\cdot \tilde{b}_{\lambda })\tilde{s}  \nonumber \\
&&+v_{0}\,\left[ (b_{\rho }^{\dagger }\cdot b_{\rho }^{\dagger }+b_{\lambda
}^{\dagger }\cdot b_{\lambda }^{\dagger })\tilde{s}\tilde{s}+s^{\dagger
}s^{\dagger }(\tilde{b}_{\rho }\cdot \tilde{b}_{\rho }+\tilde{b}_{\lambda
}\cdot \tilde{b}_{\lambda })\right]  \nonumber \\
&&+\sum_{l=0,2}c_{l}\,\left[ (b_{\rho }^{\dagger }\times b_{\rho }^{\dagger
}-b_{\lambda }^{\dagger }\times b_{\lambda }^{\dagger })^{(l)}\cdot (\tilde{b%
}_{\rho }\times \tilde{b}_{\rho }-\tilde{b}_{\lambda }\times \tilde{b}%
_{\lambda })^{(l)}+4\,(b_{\rho }^{\dagger }\times b_{\lambda }^{\dagger
})^{(l)}\cdot (\tilde{b}_{\lambda }\times \tilde{b}_{\rho })^{(l)}\right] 
\nonumber \\
&&+c_{1}\,(b_{\rho }^{\dagger }\times b_{\lambda }^{\dagger })^{(1)}\cdot (%
\tilde{b}_{\lambda }\times \tilde{b}_{\rho
})^{(1)}+\sum_{l=0,2}w_{l}\,(b_{\rho }^{\dagger }\times b_{\rho }^{\dagger
}+b_{\lambda }^{\dagger }\times b_{\lambda }^{\dagger })^{(l)}\cdot (\tilde{b%
}_{\rho }\times \tilde{b}_{\rho }+\tilde{b}_{\lambda }\times \tilde{b}%
_{\lambda })^{(l)}~,  \label{ham}
\end{eqnarray}
within the space of the totally symmetric representations $[N]$ of $U(7)$.
The coefficients $\epsilon _{s}$, $\epsilon _{p}$, $u_{0}$, $u_{1}$, $v_{0}$%
, $c_{0}$, $c_{1}$, $c_{2}$, $w_{0}$ and $w_{2}$ parametrize the
interactions. The Hamiltonian $H$ is the most general Hamitonian that
preserves angular momentum and parity, transforms as a scalar under
permutations (we consider here the case of three identical particles) and is
at most quadratic (two-body interactions). Associated with the Hamiltonian $%
H $, there are transition operators, $T$. Electromagnetic transition rates
and form factors can all be calculated by considering the matrix elements of
the operator 
\begin{eqnarray}
T &=&\mbox{e}^{-iq\beta D_{\lambda ,z}/X_{D}}~,  \nonumber \\
D_{\lambda ,z} &=&(b_{\lambda }^{\dagger }\times \tilde{s}-s^{\dagger
}\times \tilde{b}_{\lambda })_{z}^{(1)}~,
\end{eqnarray}
which is the algebraic image of the operator $\exp (iqr_{3z})$ obtained from
the full operator $\sum_{i=1}^{3}e^{i\vec{q}\cdot \vec{r}_{i}}$ by choosing
the momentum transfer $\vec{q}$ in the $z$ direction and considering
identical particles (the coefficient $X_{D}$ is a normalization factor).

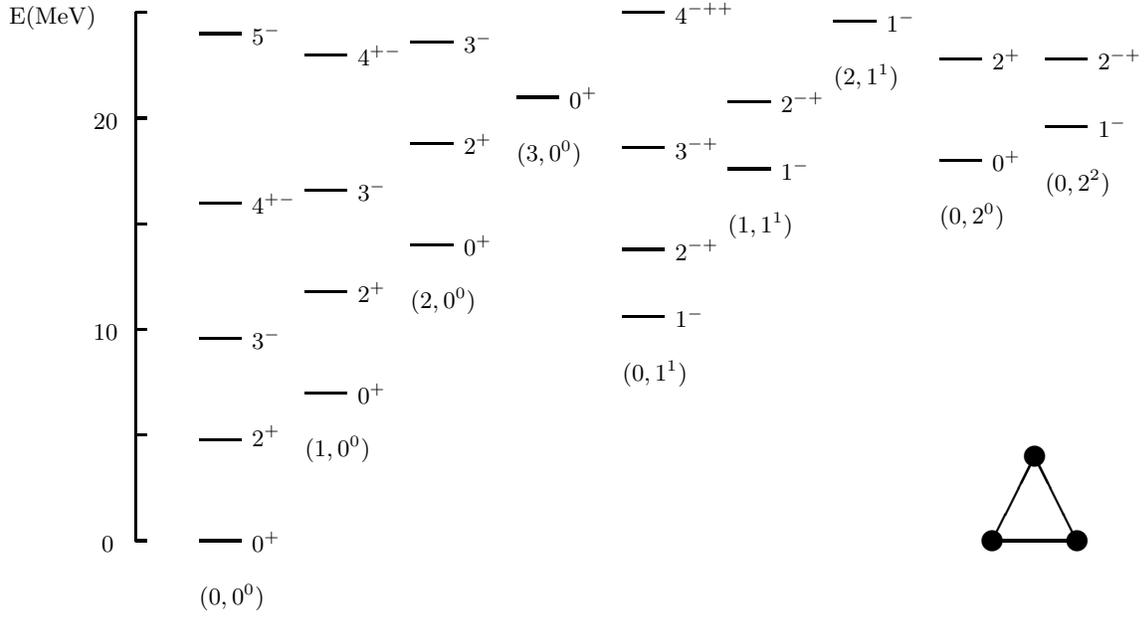
\begin{figure}
\centering
\vspace{15pt}
\setlength{\unitlength}{0.8pt}
\begin{picture}(560,350)(-60,0)
\small
\thinlines
\put (  0, 60) {\line(0,1){250}}
\thicklines
\put (  0, 60) {\line(1,0){5}}
\put (  0,110) {\line(1,0){5}}
\put (  0,160) {\line(1,0){5}}
\put (  0,210) {\line(1,0){5}}
\put (  0,260) {\line(1,0){5}}
\put (  0,310) {\line(1,0){5}}
\put (-20, 55) { 0}
\put (-20,155) {10}
\put (-20,255) {20}
\put (-60,305) {E(MeV)}
\put ( 30, 60) {\line(1,0){20}}
\put ( 30,108) {\line(1,0){20}}
\put ( 30,156) {\line(1,0){20}}
\put ( 30,220) {\line(1,0){20}}
\put ( 30,300) {\line(1,0){20}}
\thinlines
\put ( 30, 30) {$(0,0^0)$}
\put ( 55, 55) {$0^+$}
\put ( 55,105) {$2^+$}
\put ( 55,151) {$3^-$}
\put ( 55,215) {$4^{+-}$}
\put ( 55,295) {$5^-$}
\thicklines
\put ( 80,130) {\line(1,0){20}}
\put ( 80,178) {\line(1,0){20}}
\put ( 80,226) {\line(1,0){20}}
\put ( 80,290) {\line(1,0){20}}
\thinlines
\put ( 80,100) {$(1,0^0)$}
\put (105,125) {$0^+$}
\put (105,173) {$2^+$}
\put (105,221) {$3^-$}
\put (105,285) {$4^{+-}$}
\thicklines
\put (130,200) {\line(1,0){20}}
\put (130,248) {\line(1,0){20}}
\put (130,296) {\line(1,0){20}}
\thinlines
\put (130,170) {$(2,0^0)$}
\put (155,195) {$0^+$}
\put (155,243) {$2^+$}
\put (155,291) {$3^-$}
\thicklines
\put (180,270) {\line(1,0){20}}
\thinlines
\put (180,240) {$(3,0^0)$}
\put (205,265) {$0^+$}
\thicklines
\put (230,166) {\line(1,0){20}}
\put (230,198) {\line(1,0){20}}
\put (230,246) {\line(1,0){20}}
\put (230,310) {\line(1,0){20}}
\thinlines
\put (230,136) {$(0,1^1)$}
\put (255,161) {$1^-$}
\put (255,193) {$2^{-+}$}
\put (255,241) {$3^{-+}$}
\put (255,305) {$4^{-++}$}
\thicklines
\put (280,236) {\line(1,0){20}}
\put (280,268) {\line(1,0){20}}
\thinlines
\put (280,206) {$(1,1^1)$}
\put (305,231) {$1^-$}
\put (305,263) {$2^{-+}$}
\thicklines
\put (330,306) {\line(1,0){20}}
\thinlines
\put (330,276) {$(2,1^1)$}
\put (355,301) {$1^-$}
\thicklines
\put (380,240) {\line(1,0){20}}
\put (380,288) {\line(1,0){20}}
\thinlines
\put (380,210) {$(0,2^0)$}
\put (405,235) {$0^+$}
\put (405,283) {$2^+$}
\thicklines
\put (430,256) {\line(1,0){20}}
\put (430,288) {\line(1,0){20}}
\thinlines
\put (430,226) {$(0,2^2)$}
\put (455,251) {$1^-$}
\put (455,283) {$2^{-+}$}
\thicklines
\put (405, 60) {\circle*{10}}
\put (445, 60) {\circle*{10}}
\put (425,100) {\circle*{10}}
\put (405, 60) {\line ( 1,0){40}}
\put (405, 60) {\line ( 1,2){20}}
\put (445, 60) {\line (-1,2){20}}
\normalsize
\end{picture}
\vspace{15pt}
\caption[]{Spectrum of an equilateral triangle configuration 
(shown in the insert) calculated using Eq.~(\protect\ref{energy}) 
with $A=7.0$, $B=9.0$, $C=0.8$ and $D=0.0$ MeV (only the levels with 
$E \leq 25$ MeV are shown). The levels are 
characterized by angular momentum and parity $L^P$, and the 
vibrational labels $(v_1,v_2^l)$. 
Note the doubling and tripling of rotational
states. The degeneracies are removed by using a value $D\neq 0$.}
\label{top}
\end{figure}

The Hamiltonian of Eq.~(\ref{ham}) has two dynamic symmetries corresponding
to the breakings of $U(7)$ onto $U(6)$ and $SO(7)$ 
\begin{equation}
U(7)\supset \left\{ \begin{array}{c} U(6) ~, \\ 
SO(7) ~. \end{array} \right. 
\end{equation}
When the Hamiltonian contains only Casimir operators of these chains, the
eigenvalue problem can be solved in closed analytic form. The corresponding
solutions describe two situations sometimes encountered in the three body
problem: (i) six-dimensional vibrational spectra $U(6)$, and (ii) an unusual
situation which we call $\omega $-unstable or $SO(7)$ limit. Both situations
will be described in a longer publication. Here instead, as an example of
application of the algebraic method, we discuss another situation that is
appropriate to three particles at the vertices of an equilateral triangle.
The spectrum of an equilateral triangle configuration can be obtained from
the Hamiltonian of Eq.~(\ref{ham}) by setting some coefficients equal to
zero and taking specific combinations of others \cite{bij} 
\begin{eqnarray}
H &=&H_{0}+\xi _{1}\,(s^{\dagger }s^{\dagger }-b_{\rho }^{\dagger }\cdot
b_{\rho }^{\dagger }-b_{\lambda }^{\dagger }\cdot b_{\lambda }^{\dagger })(%
\tilde{s}\tilde{s}-\tilde{b}_{\rho }\cdot \tilde{b}_{\rho }-\tilde{b}%
_{\lambda }\cdot \tilde{b}_{\lambda })  \nonumber \\
&&+\xi _{2}\,\left[ (b_{\rho }^{\dagger }\cdot b_{\rho }^{\dagger
}-b_{\lambda }^{\dagger }\cdot b_{\lambda }^{\dagger })(\tilde{b}_{\rho
}\cdot \tilde{b}_{\rho }-\tilde{b}_{\lambda }\cdot \tilde{b}_{\lambda
})+4(b_{\rho }^{\dagger }\cdot b_{\lambda }^{\dagger })(\tilde{b}_{\lambda
}\cdot \tilde{b}_{\rho })\right]   \nonumber \\
&&+\xi _{3}\,(b_{\rho }^{\dagger }\tilde{b}_{\rho }+b_{\lambda }^{\dagger }%
\tilde{b}_{\lambda })^{(1)}\cdot (b_{\rho }^{\dagger }\tilde{b}_{\rho
}+b_{\lambda }^{\dagger }\tilde{b}_{\lambda })^{(1)}  \nonumber \\
&&+\xi _{4}\,(b_{\rho }^{\dagger }\tilde{b}_{\lambda }-b_{\lambda }^{\dagger
}\tilde{b}_{\rho })^{(0)}\cdot (b_{\lambda }^{\dagger }\tilde{b}_{\rho
}-b_{\rho }^{\dagger }\tilde{b}_{\lambda })^{(0)}~.  \label{ham1}
\end{eqnarray}
This spectrum does not correspond to a dynamic symmetry, since it cannot be
written in terms of invariants of a chain of algebras originating from $U(7)$%
. However, an approximate expression for the energy levels can be obtained
by making use of the method of intrinsic or coherent states (valid in the
limit of large $N$). The energy eigenvalues are then given by 
\cite{bij,BL} 
\begin{equation}
E(v_{1},v_{2}^{l},L,K,M)=E_{0}+A\,v_{1}+B\,v_{2}+C\,L(L+1)+D\,(K\pm 2l)^{2}~,
\label{energy}
\end{equation}
where $A\approx 4N\xi _{1}$, $B\approx 2N\xi _{2}$, $C=\xi _{3}/2$ and $%
D=\xi _{4}/3$. The quantum numbers have the following meaning: $v_{1}$, $%
v_{2}$ are vibrational quantum numbers; for three identical particles one of
the vibration ($v_{1}$) is singly degenerate, while the other ($v_{2}$) is
doubly degenerate; $l=v_{2},v_{2}-2,\ldots ,1$ or $0$ is the vibrational
angular momentum of the doubly degenerate vibration; $L$ is the angular
momentum, $M$ its projection on a laboratory fixed axis and $K$ its
projection on a body fixed axis. We note the particular angular momentum
composition of the rotation-vibration bands. The vibrationless ground state
band $(v_{1},v_{2}^{l})=(0,0^{0})$ has $K=3n$ ($n=0,1,2,\dots $) with $%
L=0,2,4,\ldots $ for $K=0$ and $L=K,K+1,K+2,\ldots $ for $K\neq 0$. The
parity is given by $P=(-)^{K}$. The stretching vibration $(1,0^{0})$
contains the same angular momenta $L^{P}=0^{+}$, $2^{+}$, $3^{-}$, $4^{\pm
},\ldots $, as the ground state band, while the bending vibration $(0,1^{1})$
has $K=3n+1,3n+2$ ($n=0,1,2,\ldots $) with $L=K,K+1,K+2,\ldots $. The
angular momentum content of the bending vibration is then $1^{-}$, $2^{\pm }$%
, $3^{\pm },\ldots $. Since we do not consider the excitation of the $\alpha 
$ particles themselves, the wave functions describing the relative motion
have to be symmetric. As a consequence, the relative sign in the last term
of Eq.~(\ref{energy}) is such that $|K\pm 2l|=3m$, a multiple of 3 \cite{BL}%
. (The energy formula obtained from the Hamiltonian $H$ of 
Eq.~(\ref{ham1}) contains
a Coriolis term which do not discuss here, since a detailed treatment of
this term requires the use of the full Hamiltonian of Eq.~(\ref{ham}),
rather than the simplified form of Eq.~(\ref{ham1})). In Fig.~\ref{top} 
we show the
energy spectrum corresponding to Eq.~(\ref{energy}). The importance of this
figure is the particular nature of the rotation-vibration spectrum of a
triangular configuration with ${\cal D}_{3h}$ symmetry. If a physical system
is claimed to be composed of three identical structureless particles at the
vertices of an equilateral triangle, then its spectrum {\it must} be as in
Fig.~\ref{top}. 
The algebraic framework produces this spectrum automatically by an
appropriate choice of parameters.

Another consequence of using the compact algebra $U(\nu +1)$ as a spectrum
generating algebra is that one can evaluate all observables in exact form.
For example, by taking matrix elements of the operator $T$ \ between the
eigenstates of $H$ obtained by matrix diagonalization, one can evaluate form
factors. When the Hamiltonian has a dynamic symmetry these can be derived in
closed form. Although the Hamiltonian of Eq.~(\ref{ham1}) does not
correspond to a dynamic symmetry, the form factors can still be obtained in
explicit form in the limit of large $N$. For transitions among the lowest
states they are given by 
\begin{eqnarray}
F(0_{1}^{+}\rightarrow 0_{1}^{+};q) &=&j_{0}(q\beta )~,  \nonumber \\
F(0_{1}^{+}\rightarrow 2_{1}^{+};q) &=&\frac{1}{2}\sqrt{5}\,j_{2}(q\beta )~,
\nonumber \\
F(0_{1}^{+}\rightarrow 3_{1}^{-};q) &=&-i\sqrt{\frac{35}{8}}\,j_{3}(q\beta
)~,  \nonumber \\
F(0_{1}^{+}\rightarrow 4_{1}^{+};q) &=&\frac{9}{8}\,j_{4}(q\beta )~, 
\nonumber \\
F(0_{1}^{+}\rightarrow 0_{2}^{+};q) &=&-\chi _{1}\,q\beta \,j_{1}(q\beta )~,
\nonumber \\
F(0_{1}^{+}\rightarrow 1_{1}^{-};q) &=&-i\chi _{2}\frac{1}{2}\sqrt{3}%
\,q\beta \,j_{2}(q\beta )~.  \label{ff}
\end{eqnarray}
Here $q$ is the momentum transfer and $\beta $ is the distance of the
particles from the center (the first three form factors were already given
in \cite{ino}). The last two form factors correspond to vibrational
excitations. The coefficients $\chi _{1}$ and $\chi _{2}$ are proportional
to the intrinsic matrix elements for each type of vibration ($v_{1}$ and $%
v_{2}$). Electromagnetic transition rates can be calculated from the $B(EL)$
values, which in turn can be obtained from the long wavelenght limit of the
form factors. In the case in which the constituents of the cluster are
extended objects (as in nuclei) the form factors and $B(EL)$ values can be
obtained by folding the point-like distribution with the charge distribution
(and eventually magnetic moment distribution) of the constituents. In the
case of clusters composed of $\alpha $ particles, the folding can be done in
a straightforward way, since the charge distribution of the $\alpha $
particle can be taken to a very good approximation as $\exp (-\alpha r^{2})$%
. The form factors for an extended distribution are then obtained from those
in Eq.~(\ref{ff}) by multiplying by $\exp (-q^{2}/4\alpha )$. They are a
crucial ingredient in understanding whether a cluster configuration is
present or not. When $N$ is finite (the situation encountered in nuclei) the
energy spectrum and form factors can be evaluated numerically using a
computer program written by one of us \cite{bij5}. In this case, vibrational
bands are no longer decoupled, but instead show an appreciable mixing
between them and, as a result, the spectrum is considerably distorted from
the energy formula of Eq.~(\ref{energy}).

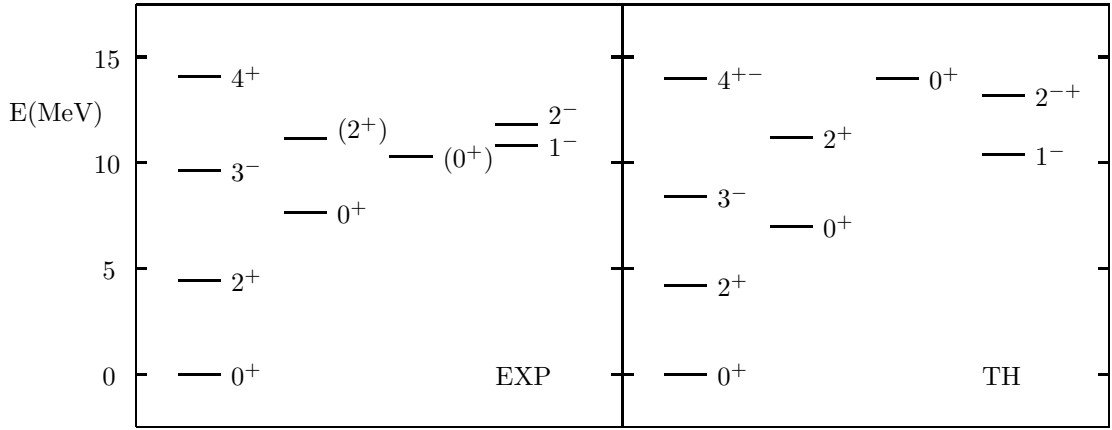
\begin{figure}
\centering
\vspace{15pt}
\setlength{\unitlength}{0.8pt}
\begin{picture}(520,250)(-60,30)
\thinlines
\put (  0, 35) {\line(0,1){200}}
\put (230, 35) {\line(0,1){200}}
\put (460, 35) {\line(0,1){200}}
\put (  0, 35) {\line(1,0){460}}
\put (  0,235) {\line(1,0){460}}
\thicklines
\put (  0, 60) {\line(1,0){5}}
\put (  0,110) {\line(1,0){5}}
\put (  0,160) {\line(1,0){5}}
\put (  0,210) {\line(1,0){5}}
\put (225, 60) {\line(1,0){10}}
\put (225,110) {\line(1,0){10}}
\put (225,160) {\line(1,0){10}}
\put (225,210) {\line(1,0){10}}
\put (455, 60) {\line(1,0){5}}
\put (455,110) {\line(1,0){5}}
\put (455,160) {\line(1,0){5}}
\put (455,210) {\line(1,0){5}}
\put (-20, 55) { 0}
\put (-20,105) { 5}
\put (-20,155) {10}
\put (-20,205) {15}
\put (-60,180) {E(MeV)}
\put (170, 55) {EXP}
\put (400, 55) {TH}
\put ( 20, 60.0) {\line(1,0){20}}
\put ( 20,104.4) {\line(1,0){20}}
\put ( 20,156.4) {\line(1,0){20}}
\put ( 20,200.8) {\line(1,0){20}}
\thinlines
\put ( 45, 55.0) {$0^+$}
\put ( 45, 99.4) {$2^+$}
\put ( 45,151.4) {$3^-$}
\put ( 45,195.8) {$4^+$}
\thicklines
\put ( 70,136.5) {\line(1,0){20}}
\put ( 70,171.6) {\line(1,0){20}}
\thinlines
\put ( 95,131.5) {$0^+$}
\put ( 95,171.6) {$(2^+)$}
\thicklines
\put (120,163.0) {\line(1,0){20}}
\thinlines
\put (145,158.0) {$(0^+)$}
\thicklines
\put (170,168.4) {\line(1,0){20}}
\put (170,178.3) {\line(1,0){20}}
\thinlines
\put (195,163.4) {$1^-$}
\put (195,178.3) {$2^-$}
\thicklines
\put (250, 60) {\line(1,0){20}}
\put (250,102) {\line(1,0){20}}
\put (250,144) {\line(1,0){20}}
\put (250,200) {\line(1,0){20}}
\thinlines
\put (275, 55) {$0^+$}
\put (275, 97) {$2^+$}
\put (275,139) {$3^-$}
\put (275,195) {$4^{+-}$}
\thicklines
\put (300,130) {\line(1,0){20}}
\put (300,172) {\line(1,0){20}}
\thinlines
\put (325,125) {$0^+$}
\put (325,167) {$2^+$}
\thicklines
\put (350,200) {\line(1,0){20}}
\thinlines
\put (375,195) {$0^+$}
\thicklines
\put (400,164) {\line(1,0){20}}
\put (400,192) {\line(1,0){20}}
\thinlines
\put (425,159) {$1^-$}
\put (425,187) {$2^{-+}$}
\end{picture}
\vspace{15pt}
\caption[]{Comparison between the low-lying experimental spectrum 
of $^{12}$C \protect\cite{ajz} and that calculated using 
Eq.~(\protect\ref{energy}) with $A=7.0$, $B=9.0$, $C=0.7$ 
and $D=0.0$ MeV. States with uncertain spin-parity 
assignment are in parentheses.}
\label{c12}
\end{figure}

The formalism introduced here can be used to study cluster states in $^{12}$%
C. It was suggested long ago \cite{whee,glass} that $^{12}$C in its ground
state can be viewed as three $\alpha $ particles at the vertices of an
equilateral triangle (point group ${\cal D}_{3h}$). The experimental
spectrum of $^{12}$C is shown in Fig.~\ref{c12}, where it is compared with that
given by Eq.~(\ref{energy}). 
One can see that this spectrum is indeed similar (if not
identical) to that of a triangular configuration. The crucial point is the
sequence of angular momenta in the ground state rotational band: $0^{+}$, $%
2^{+}$, $3^{-}$, $4^{+},\ldots $. This sequence is typical of a triangular
configuration. A linear configuration would not have negative parity states,
while a shell-model configuration would not have the $3^{-}$ state as a
member of the rotational band but rather as an octupole vibration, i.e. it
would not form a rotational sequence with the $0^{+}$, $2^{+}$, $4^{+}$
states. However, the rotational spectrum does not follow precisely what
expected from a triangular configuration (oblate top, $D<0$ in Eq.~(\ref
{energy})) but it shows rather a spherical or slightly prolate top with $%
{\cal D}_{3h}$ symmetry. The spectrum also shows an excited $0^{+}$ state at
7.65 MeV and an excited $1^{-}$ state at 10.84 MeV which could be
interpreted as bandheads of the vibrational (stretching and doubly
degenerate bending) excitations. Whether or not this is the case or rather
those states represent other types of configurations, such as three $\alpha $
particles on a line as suggested by several authors, remains an open
question. To settle this question uniquely one would have to identify the
rotational sequences built on top of them which have a characteristic
pattern for triangular configurations and another pattern for linear
configurations. In particular the nature of the $2^{+}$ state at 11.16 MeV
and $2^{-}$ state at 11.83 MeV, which could form the rotational excitation
of the doubly degenerate vibration, should be further investigated (the
role played by the $2^{+}$ state in determining the cluster structure of $%
^{12}$C has been emphasized before \cite{betts}). We have also calculated
form factors and electromagnetic transition rates \cite{tum}. All members of
the ground rotational band are well described by Eq.~(\ref{ff}), as well as
the shape of the form factors leading to the $0^{+}$ state at 7.65 MeV 
and the $1^{-}$ state at 10.84 MeV. This
analysis will be presented in a forthcoming publication \cite{tum}. The
result of the simultaneous investigation of spectra, transition rates and
form factors done within $U(7)$ is that an $\alpha $ clustering structure
(albeit not a rigid one) with ${\cal D}_{3h}$ symmetry is a good description
of the ground state configuration of $^{12}$C. However, in order to make
this conclusion stronger, we suggest to readdress the problem of $\alpha $
clustering in $^{12}$C by a remeasurement of the properties of the
high-lying states by $(\alpha ,\alpha ^{\prime })$ and $(e,e^{\prime })$
inelastic scattering. These experiments were done long ago and can benefit
from new and improved techniques. We have predictions for all form factors,
transition rates and energies of cluster states in the ${\cal D}_{3h}$
configuration. They can be obtained from us upon request.

In conclusion, we have proposed a description of cluster states in nuclei in
terms of the group $U(\nu +1)$ and shown that within this algebraic
structure one can describe many situations. In particular, for the 
three-body problem, one can recover the case of three particles at 
the vertices of a triangle, a configuration of interest in $^{12}$C. 
We have shown that $U(7)$ contains the main properties of clustering 
in nuclei: the softness of the cluster configuration, the near 
equality of vibrational and rotational energies, the spatial extension 
of the constituents and the permutation symmetry. 
We can also describe
the situation of three particles on a line (not discussed here) and of
vibrational spectra, in other words the method is flexible enough that it
can accomodate many situations encountered in nuclei. We have also
constructed the algebra appropriate to four-body problems, $U(10)$, where
additional geometric arrangements can occur, such as four particles at the
vertices of a tetrahedron (point group ${\cal T}_{d}$) and used it to study
cluster configurations in $^{16}$O. In other words, all cluster structures
up to four-body clusters can be studied with the algebraic method. The
importance of using $U(\nu +1)$ for cluster states lies in the possibility
of describing the variety of situations encountered in nuclei where clusters
are not rigid structures but rather liquid like structures arising from the
nature of the nucleon-nucleon force (spin-isospin) and the shell structure.
The unitary algebra $U(\nu +1)$ can also be of interest in the description
of other quantum mechanical systems with non-rigid structure, such as atomic
clusters, floppy molecules, and trimers making the method of broad
applicability to a large class of problems.

This work was supported in part by DGAPA-UNAM under project IN101997, 
by CONACyT under project 32416-E, and by D.O.E. Grant DE-FG02-91ER40608.

\clearpage

\end{document}